# Discernable signature of QPT in a bilayer-quantum-well system at the filling fraction ν = 5/2 in the low temperature range (1K-100 K)

Partha Goswami

Physics Department, D.B.College( University of Delhi), Kalkaji, New Delhi-110019, India
*Email id of communicating author: physicsgoswami@gmail.com

**Abstract** We consider the spin polarized fermions for the filling fraction ν = 5/2 in a bi-layer quantum well system. Since the kinetic energy of the system in fractional quantum Hall states is totally quenched, the Hamiltonian describing the system comprises of the electron correlation and tunneling terms. The correlations are captured by the 'so-called' Haldane pseudo-potentials. We employ the finite-temperature formalism involving Matsubara propagators to deal with this Hamiltonian. We show that the system undergoes a zero-order quantum phase transition (QPT), at fixed charge imbalance regulatory parameter Δ and constant layer separation as the inter-layer tunneling strength $\Delta_{SAS}$ is increased, from the effective two-component state (two independent layers) to an effective single-component state (practically a single layer). At finite and constant $\Delta_{SAS}$, a transition from the latter state to the former state is also possible upon increasing the parameter Δ. We identify the order parameter to describe this QPT as a pseudo-spin component (analogous to the component $S^z$ of the single spin-1/2 operator **S**) and calculate the order parameter with the aid of the Matsubara propagators. The clear finger-print of this QPT is obtained up to temperature equal to 100 K.

**Keywords:** Fractional quantum Hall state, Haldane pseudo-potentials, Quantum phase transition, Inter-layer tunnelling strength, Charge imbalance regulatory parameter.

PACS: 74.20.-z , 74.72.-h, 74.20.Rp, 74.72.Bk,74.20.De

We visualize a bi-layer system as the one consisting of two parallel (quasi-)two-dimensional electron systems of width 'w' separated from one another by a tunnelling barrier of thickness d (d ≥ (w/2)). The barrier height and thickness can be adjusted such that electrons are either localized in separate layers or delocalized between the two layers. We consider spin-less fermions for the filling fraction ν = 1+1+ ½ = 5/2 confined in this planar geometry. The well and the barrier materials, respectively, are assumed to be GaAs and AlGaAs. Since the kinetic energy of a system in the fractional quantum Hall states (FQHS) is totally quenched, the Hamiltonian comprises of the electron correlation (captured by the 'so-called' Haldane pseudo-potentials[1](HPP)), the inter-layer tunneling($\Delta_{SAS}$), and the charge imbalance regulatory(Δ) terms[2,3,4,5]. The Hamiltonian, we consider, is in the symmetric-anti-symmetric basis[2,3,4,5](SAS) and expressed in units of ( $e^2 /\varepsilon\, l_B$ ) where the magnetic length $l_B = \sqrt{(\hbar/eB)}$ is the length unit and ε is the permittivity of the system material. The reason for considering the $\Delta_{SAS}$ and Δ involving terms is that the recent experiments[6] have achieved bi-layer fractional quantum Hall(FQH) systems where both the inter-layer and charge imbalance tunneling terms can be controlled by changing system parameters such as gate voltages.

The second Landau level (SLL)(n =1) ( 2 < ν < 4) electron correlations are different than those in the lowest Landau level (n = 0)(LLL)( filling fraction ν = 1/2 ) as was shown by Gossard et al.[7] in their benchmark discovery of ν = 5/2 even denominator fractional quantum Hall state(FQHS). In fact, the ground state corresponding to the former is described by the non-Abelian Moore-Read Pfaffian (Pf) [8] where two particles of different types are not averse to being located at the same point in real space. The ground state corresponding to the latter, on the other hand, is known to be an Abelian Halperin 331 bilayer one[3,4]. The multiple zeros in the Halperin state, as in the Laughlin state, ensure that the amplitude of the states, for the states together to be close, always tends towards zero; in the real space two particles of different types avoid being located at the same point. The reasons for considering a bi-layer quantum well system for ν = 5/2 and not ν = ½ in this communication are (i) the effective electron-electron interaction in the ν = 5/2 case is weaker (and therefore easier to deal with analytically compared to that in the ν = ½ case) due to the spin-polarized electron (inhabiting the lowest LL) induced screening, and (ii) though the single-layer ν = 5/2 SLL FQHS, with an activation gap of order 100-500mK, has been observed by a number of experimentalists [9,10,11,12,13,14,15] in samples with high mobility($> 10^3 m^2$-$V^{-1}$-$s^{-1}$ ) at T < 100mK, the ν = ½ LLL FQHS has never been observed experimentally[15]. It may be pointed out that, whereas the odd-denominator incompressible FQH states (e.g. 1/3, 1/5, 7/3, 11/5) are robust and the even denominator (e.g. 1/2, 1/4, 5/2, 9/4) FQH states are comparatively fragile[16], never-the-less the experimental ν = 5/2 FQHE is always among the strongest observed FQH states in the latter category.

In this brief report we intend to show that, in the ν = 5/2 case, a bi-layer system undergoes a zero-order quantum phase transition (QPT), at fixed charge imbalance regulatory parameter Δ and constant layer separation as the inter-layer tunneling strength $\Delta_{SAS}$ is increased, from the effective two-component state (two independent layers) to an effective single-component state (practically a single layer). At finite and constant $\Delta_{SAS}$ a transition from the latter state to the former state is also possible upon increasing Δ. In the previous theoretical works[2,3,4,5], the former state has been linked with the Halperin Abelian 331 FQHS and the latter with the non-Abelian Moore-Read Pfaffian FQHS. It is also reported that theoretically it is not possible for the Pf FQHS to exist in the LLL, and therefore this type of phase transition is unlikely for the ν = ½ case. We wish to obtain the finger-print of the said QPT at finite but low temperatures (1K-100 K) and attempt to interpret the outcomes in the language of the thermal phase-transition. For this reason we do not opt for the zero-temperature variational wave-function approach of the previous authors. Our approach involves starting with a suitable Hamiltonian with HPPs. The Hamiltonian, in the SAS basis (or equivalently, in the right and the left layer basis), apart from the electron correlations also contains the terms accounting for the inter-layer tunneling and the charge imbalance for SLL. Under the independent LL assumption, the Hamiltonian for a LL(indexed by n = 0,1,2,…) in a compact form is given by

$$H^{(n)} = (1/2)\sum_{m,\sigma=S,AS} (d^\dagger_{m,n,S} \; d^\dagger_{m,n,AS}) \begin{bmatrix} -\Delta_{SAS} & \Delta \\ \Delta & \Delta_{SAS} \end{bmatrix} (d^\dagger_{m,n,S} \; d^\dagger_{m,n,AS})^{Transpose}$$

$$+ (1/2)\sum_{m, m', m'', m''', \sigma, \sigma'} V^{(n)}_{\{m\}} \; d^\dagger_{m,n,\sigma} d^\dagger_{m',n,\sigma'} d_{m'',n,\sigma'} d_{m''',n,\sigma} \; \delta(m + m' - m'' - m''') \quad (1)$$

where σ=(S,AS). The operators $d_{m,S}$ and $d_{m,AS}$ destroy an electron in the symmetric (S) and anti-symmetric (AS) superposition states, respectively; the index 'm' corresponds to the relative angular momentum between two electrons in a LL. The Haldane pseudo-potential(HPP) functions $V_m^{(n)}$ (which correspond to a complete set of basis functions due to angular momentum conservation)for electrons confined to a LL with index 'n', in the planar geometry, are written as $V_m^{(n)} = {}_0\int^\infty dk\ k\ [L_n(k^2/2)]^2\ L_m(k^2)\ \exp(-k^2)\ V(k)$ where $L_n(x)$ are Laguerre polynomials, and $V(k)$ is the Fourier transform of the interaction potential $V(r)$. We have $V(k) = (1/2\pi) \int d^2 r\ \exp(ik.r)\ V(r) = {}_0\int^\infty dr\ r\ J_0(k.r)\ V(r)$. Since we are considering spin polarized fermions, only odd pseudo-potentials are relevant here. Including the finite thickness effect (FTE) we notice that in the SLL (and LLL) FTE corresponds to softening all of the pseudo-potentials in a "trivial" way for $(d/l_B) < 1$ as well as $(d/l_B) \geq 1$. We have considered two finite thickness potential models[2,3,4,5], viz. (i) the infinite square-well (SQ) potential (which is appropriate for 2D GaAs quantum well structures), and (ii) the Fang-Howard (FH) variational potential (for a hetero-structure) for this purpose. In the dimensionless form, the HPPs for the former are

$$V_m^{(n)\ (SQ)} = (d/l_B)^{-1}\int dx\ [L_n((xd/l_B)^2/2)]^2\ L_m((xd/l_B)^2)\ \exp(-(xd/l_B)^2)$$

$$\times [3(xd/l_B) + (8\pi^2/(xd/l_B)) - \{32\pi^4(1-e^{-x(d/l_B)})/(x^4(d/l_B)^4 + 4\pi^2 x^2 (d/l_B)^2)\}]$$

$$\times [x^2 (d/l_B)^2 + 4\pi^2]^{-1}, \quad (2)$$

and, for the latter, the HPPs are

$$V_m^{(n)\ (FH)} = (d/l_B)^{-1}\int dx\ [L_n((xd/l_B)^2/2)]^2\ L_m((xd/l_B)^2)\ \exp(-(xd/l_B)^2)$$

$$\times (9/8)\ [(24 + 9(xd/l_B) + (xd/l_B)^2) \times (3 + (xd/l_B))^{-3}]. \quad (3)$$

The results obtained from (2) and (3) for different values of $(d/l_B)$ are summarized in Table 1. The table clearly indicates that the inequality $V_1^{(n)} > V_3^{(n)} > V_5^{(n)} > \ldots$ does not get violated for $(d/l_B) < 1$ as well as $(d/l_B) \geq 1$. In other words, due to finite thickness effect, trivial (monotonic behavior) softening of all Haldane pseudo-potentials takes place. The numerical values in Table 1 will be required in the graphical representation of the order parameter(of QPT) identified below.

We employ the finite-temperature formalism involving Matsubara propagators to deal with the Hamiltonian in (1). We identify the order parameter as a pseudo-spin component to describe the QPT and calculate the order parameter with the aid of these propagators. We shall now explain below how this identification is possible. The single spin-1/2 operator **S** is represented in terms of Pauli matrices. The basis states here are eigen states of $S_z$, i.e. $|\uparrow\rangle$, and $|\downarrow\rangle$. This operator in the second quantized language can be written as $\mathbf{S}^i = \sum_{\mu,\mu'} d^\dagger_\mu S^i_{\mu,\mu'} d_{\mu'}$ where $d^\dagger_\mu$ creates a particle in the state $|\mu\rangle$. This immediately gives $S^x = (1/2)(d^\dagger_\uparrow d_\downarrow + d^\dagger_\downarrow d_\uparrow)$, $S^y = (1/2i)(d^\dagger_\uparrow d_\downarrow - d^\dagger_\downarrow d_\uparrow)$, and $S^z = (1/2)(d^\dagger_\uparrow d_\uparrow - d^\dagger_\downarrow d_\downarrow)$. The spin-reversal operators are $S^+ = d^\dagger_\uparrow d_\downarrow$ and $S^- = d^\dagger_\downarrow d_\uparrow$. In the Hamiltonian (1) the terms $T_t = [-(\Delta_{SAS}/2)\sum_m (d^\dagger_{m,S} d_{m,S} - d^\dagger_{m,AS} d_{m,AS})]$ and $T_b = [(\Delta/2)\sum_m (d^\dagger_{m,S} d_{m,AS} + d^\dagger_{m,AS} d_{m,S})]$, respectively, account for the inter-layer tunneling and the charge imbalance. The comparison of $S^z$ and $S^x$ with the terms $T_t$ and $T_b$, respectively, shows that whereas, for a given m, $T_t^{(m)}$ may be represented by the pseudo-spin operator $S^z_m = (1/2)(d^\dagger_{m,S} d_{m,S} - d^\dagger_{m,AS} d_{m,AS})$, the operator $T_b^{(m)}$ may be represented by the pseudo-spin operator $S^x_m = (1/2)(\mathbf{S}^+_m + \mathbf{S}^-_m) = (1/2)(d^\dagger_{m,S} d_{m,AS} + d^\dagger_{m,AS} d_{m,S})$. Whereas the effect of averaging $\mathbf{S}^z_m$ over the pseudo-spin states, viz. symmetric ($|S\rangle$) and anti-symmetric ($|AS\rangle$) states [2,3,4,5], is to yield eventually the number of electrons in the symmetric and the anti-symmetric states, the effect of $S^x_m$ on the pseudo-spin states to 'switch' the states. We note that in the bi-layer problem with the total number of carriers ($2 N_A A$), where $N_A$ is the areal density and 'A' is the area of each of the layers, the total number of particles in each layer is ($N_A A$) when the electron density is balanced in each layer. Using the transformations given by $|S\rangle = (1/\sqrt{2})( |R\rangle + |L\rangle)$, and $|AS\rangle = (1/\sqrt{2})( |R\rangle - |L\rangle)$ where $|R\rangle$ and $|L\rangle$, respectively, correspond to the right and the left layer basis states, it is easy to see that the average of $S^x_m$ is non-zero for the symmetric and anti-symmetric superposition states and it is zero when there is no superposition. Thus, the operator $T_b$ represented by the pseudo-spin operator $S^x_m$ indeed accounts for the charge imbalance. In the layer basis, in which the operators $d_{m,R}$ and $d_{m,L}$ destroy an electron in the right and left quantum well respectively, we have $d_{m,S} = (d_{m,R}+d_{m,L})/\sqrt{2}$ and $d_{m,AS} = (d_{m,R} - d_{m,L})/\sqrt{2}$. In terms of $(d_{m,R}, d_{m,L})$, we have $T_t = [-(\Delta_{SAS}/2)\sum_m (d^\dagger_{m,R} d_{m,L} + d^\dagger_{m,L} d_{m,R})]$ and $T_b = [(\Delta/2)\sum_m (d^\dagger_{m,R} d_{m,R} - d^\dagger_{m,L} d_{m,L})]$. Written in this basis, while the latter indicates the possibility of transition from a practically single-layer state to an effective bi-layer one upon increasing $\Delta$ at a constant $\Delta_{SAS}$, the former indicates the possibility of transition from an effective bi-layer state to a practically single-layer one upon increasing $\Delta_{SAS}$ at a constant $\Delta$. It follows that the average of the pseudo-spin operators $\sum_m S^x_m$ or $\sum_m S^z_m$ may be chosen as the order parameter to investigate the QPT here. We have chosen the latter average to be our order parameter.

The first step of our analysis involves the calculation of (imaginary) time evolution of the operators $d_{m,n=1,\sigma}(\tau)$ where, in units such that $\hbar =1$, $d_{m,n=1,\sigma}(\tau) = \exp(H^{(1)}\tau) d_{m,1,\sigma} \exp(-H^{(1)}\tau)$. For the operator $d_{m,1,AS}(\tau)$, we obtain

$$(\partial/\partial\tau)d_{m,1,AS}(\tau) = (\mu - \Delta_{SAS})d_{m,1,AS}(\tau) - \Delta d_{m,1,S}(\tau) - \sum_{m',\sigma'} V^{(1)}_{m'} n^{(1)}_{m',\sigma'}(\tau) d_{m,1,AS}(\tau). \quad (4)$$

Here $\sigma = S/AS$, $n^{(1)}_{m',\sigma'} = d^\dagger_{m',1,\sigma'} d_{m',1,\sigma'}$, and $\mu$ is the chemical potential of the fermion number. For the operator $d_{m,1,S}(\tau)$, on the other hand, we obtain

$$(\partial/\partial\tau)d_{m,1,S}(\tau) = (\mu + \Delta_{SAS})d_{m,1,S}(\tau) - \Delta d_{m,1,AS}(\tau) - \sum_{m',\sigma'} V^{(1)}_{m'} n^{(1)}_{m',\sigma'}(\tau) d_{m,1,S}(\tau). \quad (5)$$

At this point we introduce a few thermal averages determined by $H^{(1)}$, viz. $G_{m,1,\sigma}(\tau) = -\langle T\{d_{m,1,\sigma}(\tau) d^\dagger_{m,1,\sigma}(0)\}\rangle$, $\check{D}_{m,1,-\sigma}(\tau) = -\langle T\{d_{m,1,-\sigma}(\tau) d^\dagger_{m,1,\sigma}(0)\}\rangle$, $\Gamma^{(1)}_{m',m,\sigma',\sigma}(\tau) = -\langle T\{n^{(1)}_{m',\sigma'}(\tau) d_{m,1,\sigma}(\tau) d^\dagger_{m,1,\sigma}(0)\}\rangle$, and $\Gamma^{(1)}_{m',m,\sigma',-\sigma}(\tau) = -\langle T\{n^{(1)}_{m',\sigma'}(\tau) d_{m,1,-\sigma}(\tau) d^\dagger_{m,1,\sigma}(0)\}\rangle$. As the next step, using (4) and (5), we find that the equations of motion (EOM) of these averages. The index $\sigma = +1$ and $-1$, respectively, for the case $\sigma = S$ and the case $\sigma = AS$. The third step is the calculation of the Fourier coefficients of these temperature Green's functions. We find that these coefficients are given by the equations

$$(i\omega_n + \mu + \sigma\Delta_{SAS})G_{m,1,\sigma}(i\omega_n) - \sum_{m',\sigma'} V^{(1)}_{m'} \Gamma^{(1)}_{m',m,\sigma',\sigma}(i\omega_n) - \Delta \check{D}_{m,1,-\sigma}(i\omega_n) = 1,$$

$$-\Delta(\langle d^\dagger_{m',1,\sigma'} d_{m',1,\sigma}\rangle - \langle d^\dagger_{m',1,\sigma} d_{m',1,\sigma'}\rangle) G_{m,1,\sigma}(i\omega_n) + (i\omega_n + \mu + \sigma\Delta_{SAS}$$

$$-\sum_{m'',\sigma''} V^{(1)}_{m''} \langle n^{(1)}_{m'',\sigma''}\rangle) \times \Gamma^{(1)}_{m',m,\sigma',\sigma}(i\omega_n) - \Delta\langle n^{(1)}_{m',\sigma'}\rangle \check{D}_{m,1,-\sigma}(i\omega_n) = \langle n^{(1)}_{m',\sigma'}\rangle,$$

$$-\Delta G_{m,1,\sigma}(i\omega_n) + (i\omega_n + \mu - \sigma\Delta_{SAS}) \times \check{D}_{m,1,-\sigma}(i\omega_n) - \sum_{m',\sigma'} V^{(1)}_{m'} \Gamma^{(1)}_{m',m,\sigma',-\sigma}(i\omega_n) = 0,$$

$$- \Delta \langle n^{(1)}_{m',\sigma'} \rangle G_{m,1,\sigma}(i\omega_n) - \Delta(\langle d^{\dagger}_{m',1,\sigma'} d_{m',1,\sigma} \rangle - \langle d^{\dagger}_{m',1,\sigma} d_{m',1,\sigma'} \rangle) \times \check{D}_{m,1,-\sigma}(i\omega_n)$$

$$+ (i\omega_n + \mu - \sigma\Delta_{SAS} - \sum_{m'',\sigma''} V^{(1)}_{m''} \langle n^{(1)}_{m'',\sigma''} \rangle) \times \Gamma^{(1)}_{m',m,\sigma',-\sigma}(i\omega_n) = 0. \qquad (6)$$

Here the averages $\langle n^{(1)}_{m,\sigma} \rangle$ involve single-particle excitation spectra $\varepsilon_{m,\sigma}^{(\pm)}$ and the coherence factors $a_\sigma^{(\pm)}$ to be specified shortly (see Eq.(9)). These equations, together with the usual equation to determine the chemical potential $\mu$ in terms of $(N_A A)$ and $\langle n^{(1)}_{m,\sigma} \rangle$ constitute the set of self-consistent equations to determine $(\langle n^{(1)}_{m,\sigma} \rangle, \mu)$. The equations are self-consistent, for the thermal averages $\langle n^{(1)}_{m,\sigma} \rangle$ involved determine as well as are determined by these equations. For m = 1,3,5, ......,(2N−1), i.e. N odd pseudo-potentials, Eq.(6) corresponds to 4N×2 equations involving as many unknowns. Thus, as little as solving these equations in a self-consistent manner with only three pseudo-potential calculated, viz. $V^{(1)}_1$, $V^{(1)}_3$, and $V^{(1)}_5$, becomes quite a task. Some information regarding the single-particle excitation spectrum which is expected to display interesting features due to the change in the carrier density and the involvement of the three key elements, viz. the inter-layer tunneling $\Delta_{SAS}$, the charge imbalance regulating parameter $\Delta$ and the pseudo-potentials $V^{(1)}_m$ (which depend upon $(d/l_B)$ as can be seen from Table 1), however, could be extracted comparatively painlessly if we make the drastic assumption that the pairings, or the thermal averages of the operators, are non-zero only if they correspond to the same angular momentum state and to the same $\sigma$ = S/AS. This is essentially a coarse-grained Hartree-like approximation which reduces the 4N×2 equation system to N×2 blocks involving four equations each. The system of equations now becomes tractable analytically, of course, at the cost of the quantitative accuracy. We, never-the-less, proceed further with the hope of gaining some relevant information at the qualitative level due to the change in the tunneling strengths (assuming $(d/l_B)$ fixed and greater than unity) and obtain

$$G_{m,1,\sigma}(i\omega_n) = a_\sigma^{(+)}(i\omega_n + \mu - V^{(1)}_m \langle n^{(1)}_{m,\sigma} \rangle + \Delta_0)^{-1} + a_\sigma^{(-)}(i\omega_n + \mu - V^{(1)}_m \langle n^{(1)}_{m,\sigma} \rangle - \Delta_0)^{-1}. \quad (7)$$

The coherence factors are given by $a_\sigma^{(\pm)} = \frac{1}{2}[1 \pm (\sigma\Delta_{SAS}/\Delta_0)]$ and $\Delta_0 = \sqrt{\{\Delta^2 + \Delta_{SAS}^2\}}$. For a given 'm', since $\sigma$ = + 1 and −1, respectively, for $\sigma$ = S and $\sigma$ = AS which will ensure $\langle n^{(1)}_{m,\sigma} \rangle$ being different in the two cases, the poles of $G_{m,1,\sigma}(i\omega_n)$ in (7) correspond to the quasi-particle energy states $\varepsilon_{m,S}^{(\pm)} - \mu = V^{(1)}_m \langle n^{(1)}_{m,S} \rangle \pm \Delta_0 - \mu$ and $\varepsilon_{m,AS}^{(\pm)} - \mu = V^{(1)}_m \langle n^{(1)}_{m,AS} \rangle \pm \Delta_0 - \mu$. The coarse-grained approximation made above makes it possible to construct something like a Landau Fermi liquid theory description directly in terms of low-energy quasi-particles. Here the quantity $\Delta_0$, determined by the inter-layer tunneling strength parameter $\Delta_{SAS}$ and the charge imbalance regulating parameter $\Delta$, brings about two-fold splitting of these states. For a given 'm', the symmetric-anti-symmetric(SAS)energy gap $G_m^{(1)} = (\varepsilon_{m,S}^{(\pm)} - \varepsilon_{m,AS}^{(\pm)})$ between these states is $V^{(1)}_m (\langle n^{(1)}_{m,S} \rangle - \langle n^{(1)}_{m,AS} \rangle)$. This gapped energy spectrum scenario, at a given areal density of carrier states, is possible when the chemical potential should be such that $\langle n^{(1)}_{m,S} \rangle$ and $\langle n^{(1)}_{m,AS} \rangle$ are not comparable. In other words, the pairing between the operators $d^{\dagger}_{mR}$ and $d_{mL}$ as well as that between $d^{\dagger}_{mL}$ and $d_{mR}$ are finite, for $(\langle n^{(1)}_{m,S} \rangle - \langle n^{(1)}_{m,AS} \rangle) = (\langle d^{\dagger}_{mR} d_{mL} \rangle + \langle d^{\dagger}_{mL} d_{mR} \rangle)$. The last line is nothing but the thermal average of the operator $\mathbf{S}^z_m$ where $\sum_m \langle \mathbf{S}^z_m \rangle$ is proposed to be chosen as the order parameter to investigate the QPT here. Now it is intuitive that our system is an effective single-layer, or a bi-layer for $(d/l_B)(=d\sqrt{(eB/\hbar)})$ less than one (ensuring uninhibited tunneling), or $(d/l_B)$ greater than one (ensuring inhibited tunneling) respectively. We, thus, notice that (i) small symmetric-anti-symmetric energy splitting corresponds to low tunneling and consequently to an effective bi-layer, and the opposite case to a uninhibited tunneling and an effective mono-layer, and (ii) upon decreasing $(d/l_B)$ from a moderately high value, say $(d/l_B) \sim 5$, at constant $\Delta_{SAS}$ and $\Delta$, a cross-over from a bi-layer state to a mono-layer one, in principle, is possible. Since the dimensionless pseudo-potentials for $(d/l_B) > 1$ are one to two order of magnitude less than those for $(d/l_B) < 1$ (see Table 1), this type of cross-over takes one from a weakly correlated to a strongly correlated state. We note that, to investigate this cross-over, one has to consider a wide quantum well (WQW) structure where increasing electronic density makes $(d/l_B)$ larger and vice versa. For the problem of bi-layer structure on hand, the meaningful variables to proceed with are $\Delta_{SAS}$ and $\Delta$ and (perhaps) not $(d/l_B)$. Here the total SAS gap is dependent on the pseudo-potentials and tunneling driven pairings and equal to $\sum_m V^{(1)}_m \langle S^z_m \rangle$. This prompts us to redefine our order parameter as $\Gamma = \sum_m V^{(1)}_m \langle S^z_m \rangle / \sum_m V^{(1)}_m$. We shall show below that the order parameter gets affected by the tunneling terms, when $(d/l_B)$ is held fixed at a value corresponding to the weakly correlated case $((d/l_B) > 1)$, leading to the possibility of a quantum phase transition (QPT). The clear finger-print of this transition is visible up to temperature as high as 100 K; at around 200 K the signature gets obliterated. In the strongly correlated regime $((d/l_B) < 1)$, it may not be possible to construct a similar description directly in terms of low-energy quasi-particles due to the pseudo-potentials being one to two order of magnitude stronger.

The final step is to calculate the variation in the order parameter $\Gamma$ due to the change in the tunneling terms. The thermal averages $\langle n^{(1)}_{m,\sigma} \rangle$ involved are given by $\langle n^{(1)}_{m,\sigma} \rangle = \sum_{j=\pm} a_\sigma^{(j)} (\exp \beta(\varepsilon_{m,\sigma}^{(j)} - \mu) + 1)^{-1}$, where $\beta = (k_B T)^{-1}$ and $\varepsilon_{m,\sigma}^{(\pm)} = \pm \Delta_0 + V^{(1)}_m \langle n^{(1)}_{m,\sigma} \rangle$. We note that these equations in conjunction with the equation to determine the chemical potential alluded to above, in fact, lead to values $\langle n^{(n)}_{m,\sigma} \rangle$, for the given number of carrier states $2(N_A A)$, following a tedious iterative procedure. As already mentioned above both the inter-layer and charge imbalance tunneling terms can be controlled by gate voltages, and, therefore for $(d/l_B)$ fixed, one can change $\Delta_{SAS}$ and $\Delta$ by tuning these voltages which also control the 2DEG density. The high density corresponds to small sub-band spacing (or low $\Delta_{SAS}$) and vice versa. Therefore, the system can essentially be tuned from a bi-layer like state at low $\Delta_{SAS}$ to a mono-layer one at high $\Delta_{SAS}$ for given $\Delta$ and $(d/l_B)$. The formal expression for the order parameter $\Gamma = \sum_m V^{(1)}_m (\langle n^{(1)}_{m,S} \rangle - \langle n^{(1)}_{m,AS} \rangle)/\sum_m V^{(1)}_m$ corresponding to this phase change may be written as

$$\Gamma = \{\sum_m V^{(1)}_m \tfrac{1}{2}[1 - (\Delta_{SAS}/\Delta_0)][(g_S^{(m)}\exp(\beta\Delta_0) + 1)^{-1} - (g_A^{(m)}\exp(-\beta\Delta_0) + 1)^{-1}]$$

$$+ \sum_m V^{(1)}_m \tfrac{1}{2}[1 + (\Delta_{SAS}/\Delta_0)][(g_S^{(m)}\exp(-\beta\Delta_0) + 1)^{-1} - (g_A^{(m)}\exp(\beta\Delta_0) + 1)^{-1}]\}/\sum_m V^{(1)}_m,$$

$$g_S^{(m)} = \exp[\beta(V^{(1)}_m \langle n^{(1)}_{m,S} \rangle - \mu)], \quad g_A^{(m)} = \exp[\beta(V^{(1)}_m \langle n^{(1)}_{m,AS} \rangle - \mu)]. \tag{8}$$

For the weak correlation regime $((d/l_B) = 1.5$ or greater), we find that the order parameter assumes a particularly simple form $\Gamma \approx (\Delta_{SAS}/\Delta_0) \tanh(\beta\Delta_0/2)$ for the chemical potential close to zero. The graphical representations of the order parameter are now easy to obtain. For the magnetic field $B \sim 10$ T we obtain $l_B \sim 6-8$ nm and the coulomb energy (

$e^2/\varepsilon\, l_B$) ~ 1-2 eV. We assume below the tunneling strengths to be two order of magnitude less than the coulomb energy and almost as much greater compared to the thermal energy ($k_BT$) for the 50-100 mK range. We show at zero as well as finite chemical potential that, at the temperature ~ 50 mK, the system undergoes a zero-order quantum phase transition (QPT) at fixed charge imbalance regulatory(CIR) parameter $\Delta$ and the constant layer separation, as the inter-layer tunneling(ILT) strength $\Delta_{SAS}$ is increased, from the effective two-component state (bi-layer fractional quantum Hall state(FQHS)) to an effective single-component state (single layer FQHS); at finite and constant $\Delta_{SAS}$ a transition from the latter state to the former state is also possible upon increasing $\Delta$ (see Figures1 and 2). We have been able to obtain the finger-print of this QPT at finite but low temperature (cf. the curves for 50 mK and 100 K in Figure 3) and interpret the outcomes in the language of the thermal phase-transition (TPT). For example, as the entropy is of increasing importance in TPT for determining the phase of systems with rising temperatures T, for the QPT here the term $\Gamma$ is to be accorded a similar status vis-à-vis increasing $\Delta_{SAS}$. We note that $\Gamma$ corresponds to an analytically tractable quantity which is found to be an increasing function of $\Delta_{SAS}$ for the given Coulomb repulsions (see Figure 3). The counterpart of the specific heat capacity here is the susceptibility $\chi \equiv (\delta\Gamma/\delta\Delta_{SAS})$. Analogous to the second-order thermal phase transition, we conclude from Fig.3 that there is a discontinuity in $\chi$ while the system undergoes transition from the bi-layer FQHS to the single-layer FQHS. It must be emphasized that unveiling of these features have been possible as we did not opt for the zero-temperature wave-function approach.

In conclusion, the semi-conductor bi-layers with finite single-layer width can support both the effective two-component state (two independent layers) and the effective single-component state (practically a single layer). In the previous theoretical works(see the zero-temperature wave-function approach in refs. [2,3,4,5] ), the former state has been linked with the Halperin Abelian 331 FQHS and the latter with the non-Abelian Moore-Read Pfaffian FQHS for the second Landau level. Also, there could be a QPT, as a function of tunneling strengths (at constant layer separation), between these states. The present work is (perhaps) a first attempt to investigate the even denominator fractional quantum Hall (FQH) effect related phenomena, in the context of the bi-layer quantum well, within the framework of finite temperature quantum field theoretic formalism. The work unveils one of the precise roles of finite (but low) temperature which is not possible to decipher in the usual zero-temperature variational wave function approach of previous authors [2,3,4,5].To elaborate, the graphical representation in Fig. 3 shows that the finger-print of the discontinuity in the susceptibility(related to the order parameter and the inter-layer tunneling strength) at quantum phase transition(QPT) in the milli-Kelvin range remains up to about 100 K and disappears as the temperature is raised further. Thus the new result, apart from the fact that the effective mono-layer state is slightly more robust compared to an effective bi-layer one as the former exists over a larger range of the tunneling strengths (see Figs. 1 and 2) compared to the latter one, is that the clear signature of a QPT in the bi-layer system is available up to liquid nitrogen temperature. This is an experimentally verifiable prediction of our finite-temperature approach. The result is significant as the serious effort is currently underway in several laboratories to lower the electron/bath temperature to few mK in order to boost the hope for topological quantum computation using non-Abelian FQH state. Our result concerning SAS gap shows that the topological protection ( larger the gap separating the many-body degenerate ground states from the low-lying excited states the more robust is the topological protection) in the non-Abelian FQHS is effective up to liquid nitrogen temperature. Therefore, it seems possible that for the non-Abelian state, in stead of adiabatic demagnetization (in dilution refrigerators to achieve tens of mK temperature), liquid nitrogen will be adequate to ensure the said protection.

**TABLES**

**Table 1** The values of the Haldane pseudo-potentials are summarized here considering **(a)** the infinite square-well (SQ) potential, and **(b)** the Fang-Howard (FH) variational potential in Eqs. (2) and (3). Since $V_1^{(n)} > V_3^{(n)} > V_5^{(n)} > \ldots$ for a given $(d/l_B)$ does not get violated, the conclusion is that finite thickness effect (FTE) corresponds to softening of all of the pseudo-potentials in a rather "trivial" way for all $(d/l_B)$. We have taken the limits of integration in Eqs.(2) and (3) as 0.001 and 10.

**(a)**

|  | $(d/l_B)$ =0.2 | $(d/l_B)$ =0.5 | $(d/l_B)$ =0.6 | $(d/l_B)$ =0.8 | $(d/l_B)$ =1.0 | $(d/l_B)$ =1.5 |
|---|---|---|---|---|---|---|
| $V_1^0$ | **11.2091** | **1.7505** | **1.2156** | **0.6838** | **0.4376** | 0.1945 |
| $V_3^0$ | **6.7849** | **1.1043** | **0.7669** | **0.4314** | **0.2761** | 0.1227 |
| $V_5^0$ | **5.5785** | **0.8709** | **0.6048** | **0.3402** | **0.2177** | 0.0968 |
| $V_1^1$ | **10.9555** | **1.6372** | **1.1370** | **0.6396** | **0.4093** | 0.1819 |
| $V_3^1$ | **7.9503** | **1.2432** | **0.8634** | **0.4857** | **0.3108** | 0.1382 |
| $V_5^1$ | **5.9444** | **0.9256** | **0.6428** | **0.3616** | 0.2314 | 0.1029 |

**(b)**

|     | $(d/l_B)=0.2$ | $(d/l_B)=0.5$ | $(d/l_B)=0.6$ | $(d/l_B)=0.8$ | $(d/l_B)=1.0$ | $(d/l_B)=1.5$ |
|---|---|---|---|---|---|---|
| $V_1^0$ | 10.4314 | 1.6441 | 1.1418 | 0.6423 | 0.4110 | 0.1827 |
| $V_3^0$ | 6.6646 | 1.0777 | 0.7484 | 0.4210 | 0.2694 | 0.1198 |
| $V_5^0$ | 5.4421 | 0.8580 | 0.5959 | 0.3352 | 0.2145 | 0.0954 |
| $V_1^1$ | 9.9730 | 1.5300 | 1.0625 | 0.5977 | 0.3825 | 0.1700 |
| $V_3^1$ | 7.4312 | 1.1763 | 0.8169 | 0.4595 | 0.2941 | 0.1307 |
| $V_5^1$ | 5.7489 | 0.9038 | 0.6277 | 0.3531 | 0.2260 | 0.1004 |

# FIGURES

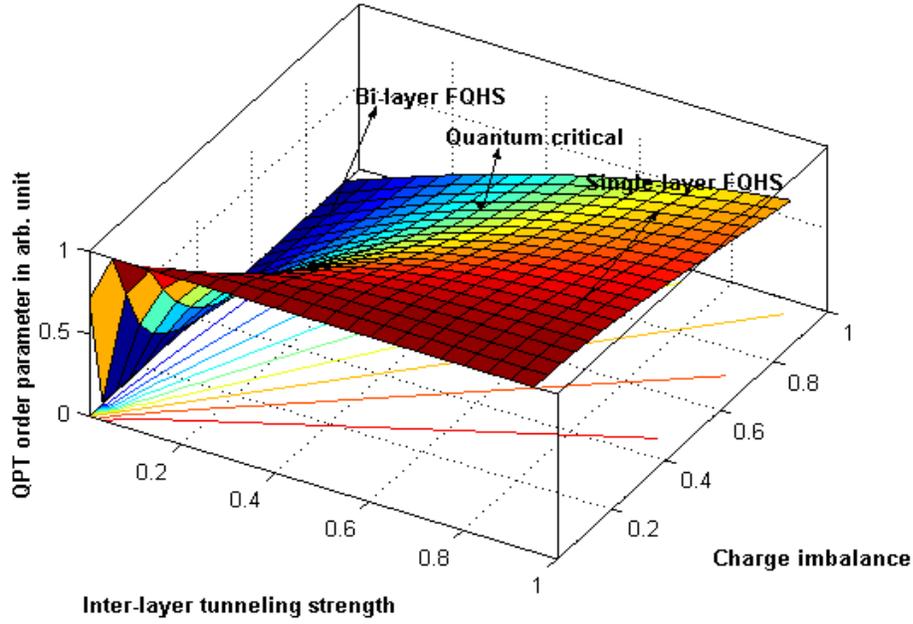

**Figure 1.** A 3-D plot of the QPT order parameter on the $\Delta_{SAS}$ - $\Delta$ zone at the temperature T = 50 mK for the spin polarized fermions at the filling fraction $\nu$ = 5/2 for zero chemical potential of fermion number. We show that, in the $\nu$ = 5/2 case, a bi-layer system undergoes a zero-order quantum phase transition (QPT), at fixed charge imbalance regulatory parameter $\Delta$ and constant layer separation as the inter-layer tunneling strength $\Delta_{SAS}$ is increased, from the effective two-component state (two independent layers) to an effective single-component state (practically a single layer). At finite and constant $\Delta_{SAS}$ a transition from the latter state to the former state is also possible upon increasing $\Delta$. In the previous theoretical works[2,3,4,5], the former state has been linked with the Halperin Abelian 331 FQHS(see Figure 4) and the latter with the non-Abelian Moore-Read Pfaffian FQHS (see Figure 5). It is also reported that theoretically it is not possible for the Pf FQHS to exist in the LLL, and therefore this type of phase transition is unlikely for the $\nu$ = ½ case.

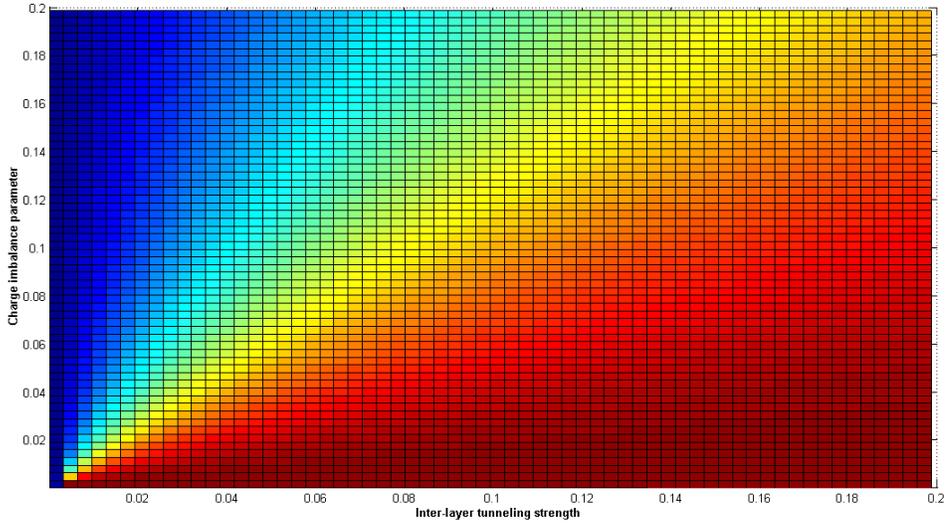

**Figure 2**. The contour plot of QPT order parameter in arbitrary unit as a function of the interlayer tunneling strength and the charge imbalance regulatory parameter for a high mobility bi-layer system at T ~ 50 mK for finite chemical potential. The cold region corresponds to bi-layer FQHS( abelian Halperin 331 FQHS) and the hot region to the single-layer FQHS(non-abelian Moore-Read Pffaian). The greenish intermediate region is for the quantum criticality. The scale of the plot is from 0 to 1.

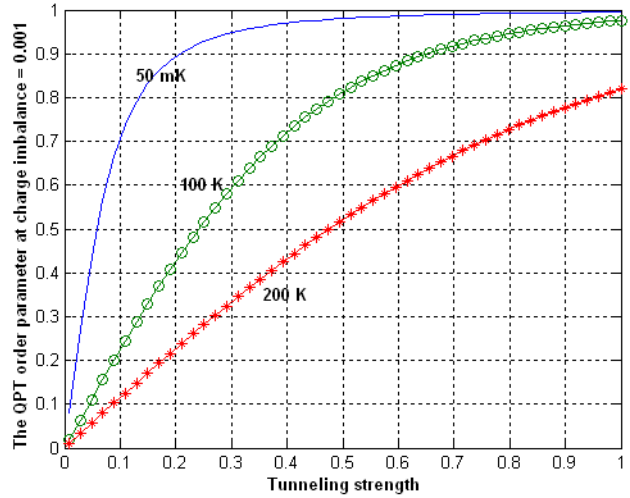

**Figure 3.** A 2-D plot of the QPT order parameter $\Gamma$ as a function of $\Delta_{SAS}$ for $\Delta = 0.001$ at the temperatures T = 50 mK, 100 K and 200 K for the spin polarized fermions at the filling fraction $\nu = 5/2$ in the weak correlation case $(d/l_B) = 1.5$. The softening of all of the pseudo-potentials occurs for $(d/l_B) > 1$. We find that there is a discontinuity in the susceptibility $\chi \equiv (\delta\Gamma/ \delta\Delta_{SAS})$ while the system undergoes transition from the bi-layer FQHS to the single-layer FQHS. Whereas the plot at 100 K carries the finger-print of QPT mentioned above, at T = 200 K the finger-print is completely obliterated. The plot indicates that the quanntum Hall effect related phenomena are strongly affected by the increase in temperature.

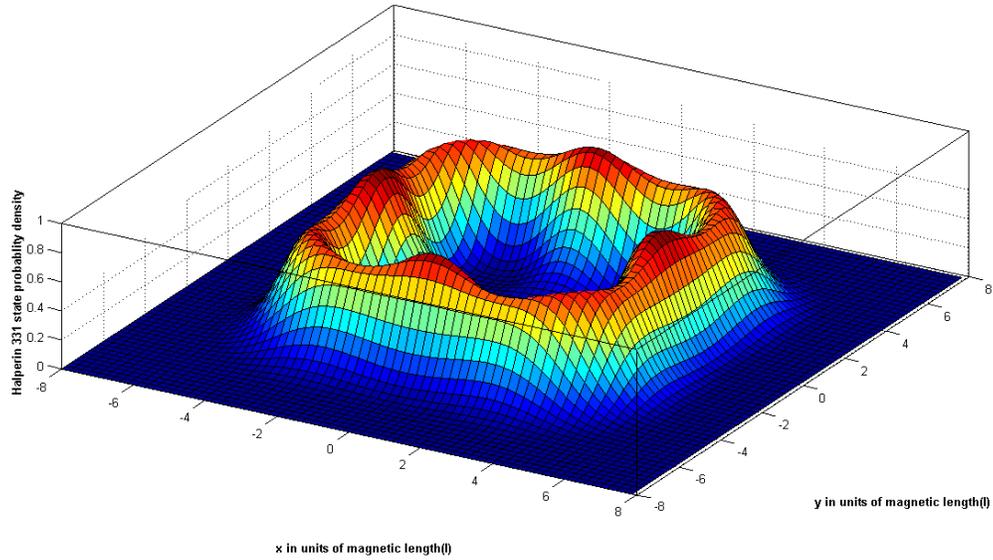

(a)

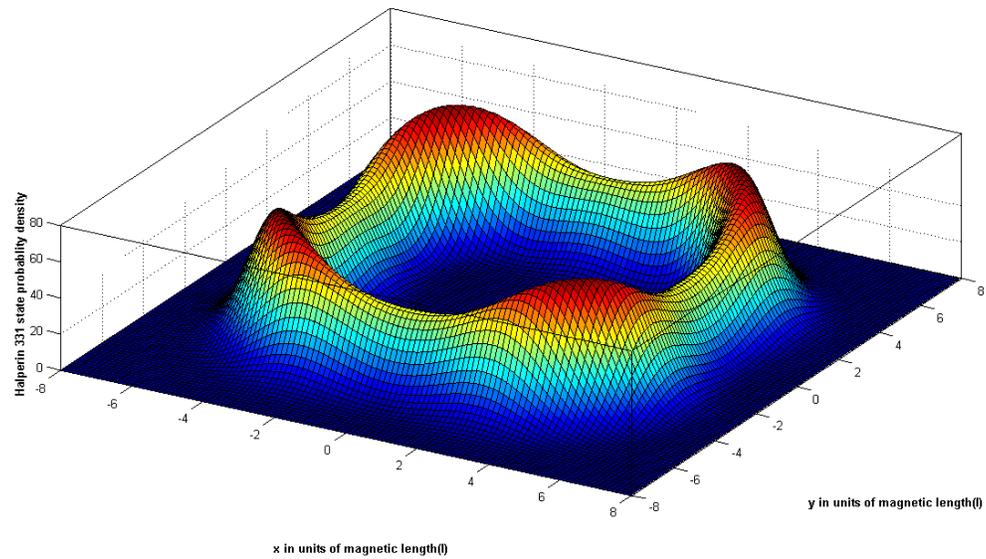

(b)

**Figure 4**. The 3-D plots of 331 Halperin (two-components a = 1,2 of five particles each ) probability density corresponding to a ten-particle system on the $L_x$-$L_y$ plane. Here $-8 \leq L_x, L_y \leq +8$. (a) We have plotted here the possible locations of the fifth particle of a = 1 type with four electrons of a = 1 type at the fixed locations, viz. at the points (0,±8) and (±8,0). The a =2 type particle locations are at (±2, ±2) and (0,0). (b) We have plotted here the possible locations of the fifth particle of a = 2 type; the four a = 2 type particle locations are at (±2, ±2). The five particles of a = 1 type are located at (±8, ±8) and (0,0). The quantum number of the electrons a = 1,2 could describe spin (up or down), layer (Right or Left), sub-band (Symmetric or Anti-symmetric). It may be noted that the fifth particle in both the cases scrupulously avoids the remaining particles in the assembly.

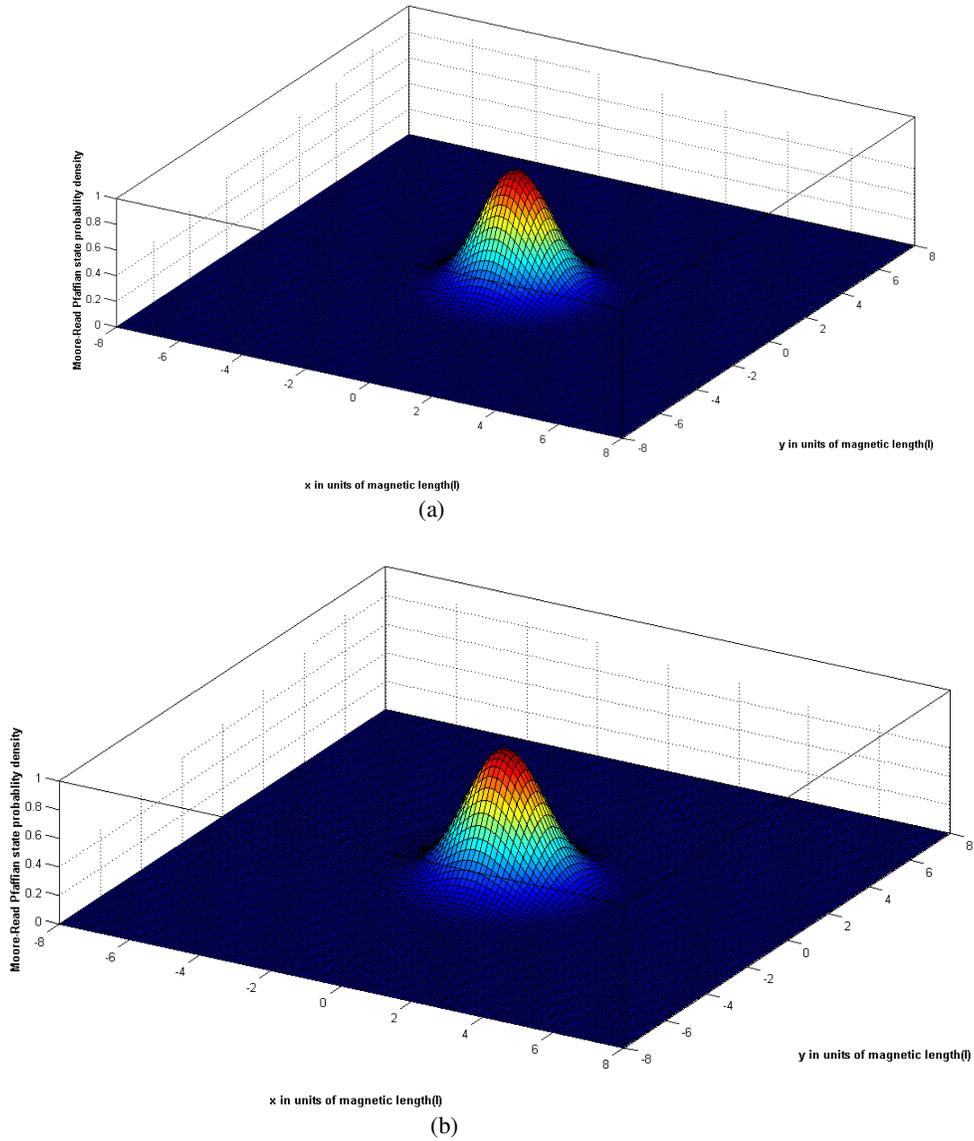

**Figure 5**. The 3-D plots of the Moore-Read Pfaffian (single-component) probability density corresponding to a ten-particle system with a = 1,2 of five particles each on the $L_x$-$L_y$ plane ($-8 \leq L_x, L_y \leq +8$). (a) We have plotted here the possible locations of the fifth particle of a = 1 type. Four electrons of a = 1 type are at the fixed locations, viz. at the points $(0, \pm 8)$ and $(\pm 8, 0)$ of the $L_x$-$L_y$ zone. The a =2 type particle locations are at $(\pm 2, \pm 2)$ and $(0,0)$. (b) We have plotted here the possible locations of the fifth particle of a = 2 type; the four a = 2 type particle locations are at $(\pm 2, \pm 2)$. The five particles of a = 1 type are located at $(\pm 8, \pm 8)$ and $(0,0)$. It may be noted that the fifth particle of a certain type, in both the cases, no more avoids a particle of the different type in the assembly.